\documentclass[a4paper,12pt]{article}
\usepackage{amsmath, amssymb}
\usepackage[hmargin=2cm,vmargin=3cm]{geometry}
\usepackage[pdftex]{graphicx}

\newcommand*{\ddp}{{\frac{\partial}{\partial p}}}
\newcommand*{\ddq}{{\frac{\partial}{\partial q}}}
\newcommand*{\ddpp}{{\frac{\partial^2}{\partial p^2}}}
\newcommand*{\bhr}{{\beta\hbar\gamma}}
\newcommand*{\mhr}{{m\hbar\gamma}}
\newcommand*{\jk}{{j_1,\cdots,j_K}}
\newcommand*{\jkp}{{j_1,\cdots,j_k+1,\cdots,j_K}}
\newcommand*{\jkm}{{j_1,\cdots,j_k-1,\cdots,j_K}}

\begin{document}
\noindent{ \Large{Quantum Suppression of Ratchet Rectification in a Brownian System Driven by a Biharmonic Force}}
\par\bigskip
\noindent{Akihito Kato${}^1$ and Yoshitaka Tanimura
\footnote{Department of Chemistry, Graduate School of Science, Kyoto University, Kyoto606-8502, Japan}
\footnote{Universit\"{a}t Augsburg, Institut f\"{u}r Physik, Universit\"{a}tsstrasse 1, 86135 Augsburg, Germany \\ email address: tanimura@kuchem.kyoto-u.ac.jp}}

\begin{abstract}
We rigorously investigate the quantum dissipative dynamics of a ratchet system
described by a periodic potential model based on the Caldeira-Leggett Hamiltonian with a biharmonic force.
In this model, we use the reduced hierarchy equations of motion in the Wigner space representation.
These equations represent a generalization of the Gaussian-Markovian quantum Fokker-Planck equation 
introduced by Tanimura and Wolynes (1991), 
which was formulated to study non-Markovian and non-perturbative thermal effects at finite temperature.
This formalism allows us to treat both the classical limit and the tunneling regimes,
and it is helpful for identifying purely quantum mechanical effects through the time evolution of the Wigner distribution.
We carried out extensive calculations of the classical and quantum currents
for various temperatures, coupling strengths, and barrier heights.
Our results reveal that at low temperature,
while the quantum current is larger than the classical current in the case of a high barrier,
the opposite is true in the case of a low barrier.
We find that this behavior results from the fact that the tunneling enhances the current in the case of a high barrier,
while it suppresses the current in the case of a low barrier. 
This is because the effect of the ratchet potential is weak in the case of a low barrier, due to the large dispersion of the distribution introduced by tunneling. This causes the spatio-temporal asymmetry, which is necessary for ratchet current, to be weak, and as a result, the net current is suppressed.
\end{abstract}

\section{Introduction}
A system that is able to rectify thermal or mechanical fluctuations through a periodic potential is called a ratchet.\cite{Smoluchowski12,Feynman66,Hanggi96LNP,Ajdari97,Astumian97,Reimann02PR,Hanggi05AnnPhys,Hanggi09}
With recent advances in microscopy and nano-technology,
several new mechanisms of rectification of fluctuations 
resulting in transport have been identified through theoretical works\cite{Magnasco93,Hanggi94EPL,Pollak98EPL,Nitzan02, Scheidl02, Seeger66,Breymayer81,Breymayer82,Breymayer84,Marchesoni86,Hanggi98EPL} 
and experimental works\cite{Linke98EPL,Linke99,Linke02AppPhys,Majer03,Villegas03, Ustinov04,
Shiavoni03,Renzoni05,Salger09,Astumian98EurBiophys, Nishiyama03} in biology, physics, and chemistry.
Well-known examples include asymmetric quantum dots\cite{Linke98EPL,Linke99,Linke02AppPhys}, 
vortices in superconductors\cite{Majer03,Villegas03, Ustinov04}, 
cold atoms in asymmetric optical lattices\cite{Shiavoni03,Renzoni05,Salger09}, 
and molecular motors in biological systems.\cite{Astumian98EurBiophys, Nishiyama03,Fisher07}
A ratchet system can produce a directed current from fluctuations only when the system is in a non-equilibrium state,
because the extraction of work from unbiased fluctuations is not allowed by the second law of thermodynamics.
Moreover, in order to realize the ratchet effects,
it is necessary to break the spatio-temporal symmetries,
because otherwise the contributions from positive and negative currents would cancel.
\cite{FlachPRL00,FlachERL01,ZolotaryukPRE02,ReimannPRL01,Quintero10}
Quantum mechanical effects may also play a role,
because ratchet systems must be of microscopic size to rectify the fluctuating motion.\cite{Hanggi97PRL,Reimann98,Hanggi98PRL,Hanggi98PRL2,Majer02,Zueco04,AnkerholdPRE10,HanggiPRA11,FlachPRA07,Hanggi09EPL,Grossmann11} 

A Fokker-Planck approach\cite{Hanggi09AJP} and a Langevin approach\cite{Machura10PRE} have been used
in classical studies of ratchets,
while varieties of equation of motion approaches have been employed for quantum mechanical studies.
\cite{Hanggi97PRL,Reimann98,Hanggi98PRL,Hanggi98PRL2,Majer02,Zueco04,AnkerholdPRE10,HanggiPRA11,FlachPRA07, Hanggi09EPL,Grossmann11}
In quantum mechanical case,
dissipative systems are commonly modeled as potential systems coupled to heat-bath degrees of freedom at finite temperature.
This coupling gives rise to thermal fluctuations and dissipation
that drive the systems toward the thermal equilibrium state.
The heat-bath degrees of freedom are then reduced
using such methods as the projection operator method and the path integral method, for example.
The quantum Langevin equation\cite{Hanggi97PRL} and the quantum Fokker-Planck equation\cite{Zueco04} have been used
for the purpose of understanding the quantum aspects of ratchet dynamics in a heat bath.
Although these equations are analogous to the classical kinetic equations,
which have proved to be useful in the treatment of classical transport problems,
such equations cannot be derived in a quantum mechanical framework without significant approximations and/or assumptions.
For example, in dealing with the quantum Langevin equation expressed in operator form,
it is generally assumed that the anti-symmetric correlation function of the noise is positive,
but this is valid only for slow non-Markovian modulation at high temperature, as we demonstrate in Sec. 2.
Similarly, the quantum Fokker-Planck equation can be derived from the Caldeira-Leggett Hamiltonian under a Markovian approximation,
\cite{CLPhysica1983,Waxman1985,Coffey1,Coffey2,Jyoti2011} but in order for this to be possible,
the heat bath must be at a sufficiently high temperature,
in which case quantum tunneling processes play a minor role.
The quantum master equation expressed in terms of Floquet states provides
a more rigorous treatment of quantum dissipative dynamics than the methods mentioned above,\cite{Hanggi09EPL}
but it can only be applied to systems possessing weak interactions.\cite{KDH1997}
The methods discussed to this point represent the main approaches
used in the treatment of ratchet systems carried out to this time.
However, as we have pointed out,
none of them provide fully quantum mechanical descriptions of broad validity.

In this paper, we employ the reduced hierarchy equations of motion (HEOM) in the Wigner space representation,\cite{TJPSJ2006} 
which are a generalization of the Gaussian-Markovian quantum Fokker-Planck equation introduced by Tanimura and Wolynes,\cite{TanimuraWolynes1,TanimuraWolynes2} 
to study non-Markovian and non-perturbative thermal effects at finite temperature.\cite{YT_Steffen, Kato_1,STJPCA2011,STJPSJ2013,STNJP2013}
Because the HEOM are derived from the full system-bath Hamiltonian with no approximation,
the entire system  approaches a 
thermal equilibrium state at finite temperature when no external perturbation is applied. \cite{TJPSJ2006,TKJPSJ1989}
This means that the second law of thermodynamics holds within the HEOM approach
and hence there can be no finite current in the absence of a driving force,
which is the essential behavior to investigate the ratchet rectification of thermal fluctuations. 
The HEOM have been used to study a variety of phenomena and systems,
including exciton transfer\cite{IshiFlem09, shi11,Kramer12,Schulten12, DijkstraNJP10},
electron transfer\cite{Xu07,Tanaka2009JPSJ,Tanaka2010JCP,TJCP2012},
quantum information\cite{Dijkstra10,Nori12},
and resonant tunneling diodes.\cite{STJPSJ2013,STNJP2013}
The HEOM are ideal for studying quantum transport systems when implemented using the Wigner representation,
because they allow the treatment of continuous systems
utilizing open boundary conditions and periodic boundary conditions\cite{Frensley}.
Elucidation of the dynamical behavior of the system
through the time evolution of the Wigner distribution functions is also possible.
The classical hierarchy equations of motion can be obtained easily by taking the classical limit of the HEOM and, 
utilizing the Wigner representation,
we can easily compare quantum and classical distribution functions.\cite{TanimuraWolynes2, STJPCA2011}
In this paper, considering both classical and quantum mechanical cases,
we report the results of numerical calculations of the ratchet current,
which is the induced net current resulting from ratchet effects,  
in a dissipative environment for various temperatures and system-bath coupling strengths.
We then clarify the roles of fluctuations, dissipation and the biharmonic force in both the classical and quantum cases. 

The organization of the paper is as follows.
In Sec. 2, we introduce the HEOM approach in the Wigner representation. 
In Sec. 3, we discuss ratchet systems from the point of view of space-time symmetries. 
In Sec. 4, the numerical results obtained for the ratchet current in classical and quantum mechanical cases are presented. 
Section. 5 is devoted to conclusions.

\section{Reduced Hierarchy Equations of Motion Formalism}
Ratchet systems are often modeled by a Brownian particle in a periodic potential under an ac or dc driving force. 
We take this approach and employ a model based on the Caldeira-Leggett (or Brownian) Hamiltonian\cite{CLAnnlPhys1983},
\begin{equation}
 \hat{H}_{tot} = \frac{{{{\hat p}^2}}}{{2m}} + U(\hat q;\;t) +
	 \sum_j \left[ \frac{\hat{p}_j^2}{2m_j} + \frac{m_j\omega_j^2}{2}\left( \hat{x}_j - \frac{a_jV(\hat{q})}{m_j\omega_j^2}\right)^2 \right]. 
\label{tot_H}
\end{equation}
Here, $m$, $\hat{p}$, $\hat q$ and $U(\hat{q},t)$ are the mass, momentum, position and potential of 
the particle, and $m_j$, $\hat{p}_j$, $\hat x_j$ and $\omega_j$ are the mass, momentum, 
position and frequency variables of the $j$th bath oscillator mode.
The quantities $a_j$ are coefficients that depend on the nature of the system-bath coupling.
From eq (\ref{tot_H}),
it is seen that the interaction part of the Hamiltonian
is assumed to take the form $\hat{H}_I = -V(\hat{q})\hat{X}$,
where $V(\hat{q})$ is an arbitrary function of $\hat{q}$ and $\hat{X}\equiv \sum_j a_j\hat{x}_j$ is the interaction coordinate.
We introduced the counter term $\sum_j a_j^2\hat{q}^2/2m_j\omega_j^2$ 
to maintain the translational symmetry of the Hamiltonian for $U(\hat{q}; 0)=0$.
Here, we consider a reflection-symmetric potential system driven by a biharmonic force.
\cite{Seeger66,Breymayer81,Breymayer82,Breymayer84,Marchesoni86,Hanggi98EPL,FlachPRA07,Hanggi09EPL,Grossmann11,Hanggi09AJP,Machura10PRE}
This force prevents the system from reaching equilibrium and breaks the spatio-temporal symmetry. 
We write the potential as $U(\hat q;\;t) = U(\hat q) -\hat{q}F(t)$,
where $U(\hat q)$ is static, reflection-symmetric potential,
and $-\hat{q}F(t)$ is the dynamic potential. Note that the time average of $F(t)$ is zero.
The total system obeys the von-Neumann equation (the quantum Liouville equation).

After the bath degrees of freedom are traced out,
the reduced density matrix elements of the system are obtained in the path integral form as
\begin{align}
 \rho(q,q';t) =& \int dq_0dq'_0\int Dq\int Dq' \rho(q_0,q'_0) 
\rho _{CS} (q,\,q',\,t;\,q_0 ,q'_0 ) \nonumber \\
	&\times \exp (iS_A[q,t]/\hbar)F[q,q';t] \exp( -iS_A[q',t]/\hbar), 
\label{rho_red}
\end{align}
where $S_A[q;t]$ is the action corresponding to the system's Hamiltonian, $\hat{H}_A={\hat p}^2/2m+U({\hat q};t)$,
$\rho (q_0 ,q'_0 )$ is the initial state of the system at time $t_0$,
and $\rho _{CS}(q, \,q', \,t; \,q_0, \,q'_0 )$ is the initial correlation function between the system and the heat bath \cite{Grabert}.

The effect of the bath is incorporated into the Feynman-Vernon influence functional,
which is written
\begin{align}
F[q,\,q';\,t] ={\rm exp}\left\{ - \frac{1}{\hbar^2}  \int_{t_0}^t {ds } V^{\times}(s) 
 \left[ -\frac{\partial }{\partial s }\int_{t_0}^{s} {du } \,\frac{i\hbar}{2} \Psi (s - u)
V^{\circ}(u) 
  + \int_{t_0}^{s} {du} C(s  - u )
V^{\times}(u)  \right] \right\}, 
\label{if}
\end{align}
where $V^\times(t) \equiv V(q(t)) - V(q'(t))$ and $V^\circ(t) \equiv V(q(t))+V(q'(t))$.
Although the method we employ here can be used with any form of $V(q)$,\cite{YT_Steffen, Kato_1,STJPCA2011}
in this paper, we consider the linear-linear system bath coupling case, defined by $V(q)=q$.
The interaction coordinate $\hat{X}\equiv \sum_j a_j\hat{x}_j$ is regarded as a driving force through the interaction $-\hat{q}\hat{X}$.
The canonical and symmetrized correlation functions, respectively, are then expressed as
$\Psi(t) \equiv \beta \langle \hat{X} ; \hat{X}(t) \rangle_\mathrm{B}$ and $C(t) \equiv \frac{1}{2} \langle \{ \hat{X}(t), \hat{X}(0) \} \rangle_{\mathrm{B}}$,
where $\beta=1/k_B T$ is the inverse temperature,
$\hat{X}(t)$ is $\hat{X}$ in the Heisenberg representation,
and $\langle \cdots \rangle_{\mathrm{B}}$ represents the thermal average over the bath modes.\cite{TKJPSJ1989,TJPSJ2006}
Using the spectral density $J(\omega)=\sum_j a_j^2\delta(\omega-\omega_j)/2m_j\omega_j $, we can rewrite these functions as
\begin{equation}
\Psi(t) = 2\int_0^{\infty}  d\omega \frac{J(\omega)}{\omega} \cos(\omega t),
\end{equation}
and
\begin{equation}
C(t) = \hbar \int_0^{\infty}  d\omega J(\omega) \cos(\omega t) \coth \left( \frac{\beta \hbar \omega}{2} \right).
\end{equation}
The function $C(t)$ is analogous to the classical correlation function of $X(t)$
and corresponds to the correlation function of the bath-induced noise (fluctuations), whereas $\Psi(t)$ corresponds to dissipation.
The function $C(t)$ is related to $\Psi(t)$ through the quantum version of the fluctuation-dissipation theorem,
$C[\omega] = \hbar \omega \coth(\beta \hbar \omega/2) /2 \Psi[\omega]$,
which insures that the system evolves toward the thermal equilibrium state for finite temperatures, $\rm{tr_B} \{ \exp[-\beta \hat{H}_{tot}] \}$,
in the case that there is no driving force\cite{KuboToda85}. 

\begin{figure}
\includegraphics[width=15cm]{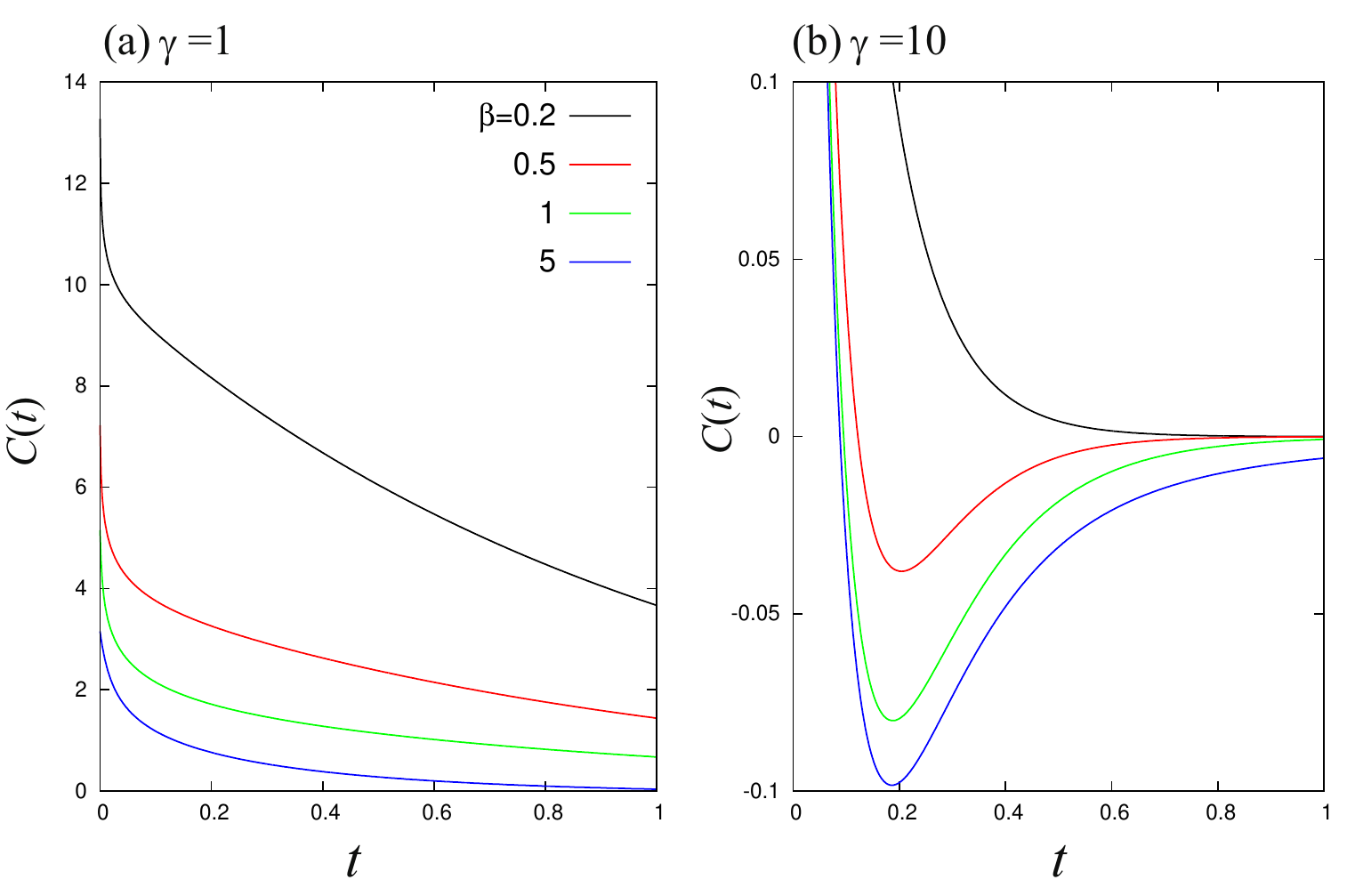}
\caption{The symmetric correlation $C(t)$ defined by eq  (\ref{eq:C}) is depicted as a function of the dimensionless time $t$ for (a) the slow modulation case, $\gamma =1$,
and (b) the fast modulation case, $\gamma=10$.
The constants $m \zeta \gamma ^2/2$ and $\hbar$ are set to unity.
Note that $\gamma \to \infty$ corresponds to the Ohmic (Markovian) limit,
as can be seen from eq (\ref{Jform}).
In each of the figures, the inverse temperatures are, from top to bottom, $\beta =0.2, 0.5, 1.0$, and $5$.
The noise correlation $C(t)$ becomes negative for (b), the fast modulation case, at low temperature (large $\beta$)
due to the contribution of Matsubara frequency terms. }
\label{C}
\end{figure}

As shown in Figure \ref{C},
the noise correlation $C(t)$ becomes negative at low temperature,
due to the contribution of the Matsubara frequency terms with $\nu=2\pi/\beta\hbar$ in the region of small $t$.
This behavior is characteristic of quantum noise.
The fact that the noise correlation takes negative values introduces problems
when the quantum Langevin equation is applied to quantum tunneling at low temperature.
We note that the characteristic time scale over which we have $C(t) < 0$ is determined by the temperature
and is not influenced by the spectral distribution $J(\omega)$.
Thus, the validity of the Markovian (or $\delta (t)$-correlated) noise assumption is limited in the quantum case to the high temperature regime.
Approaches employing the Markovian master equation and the Redfield equation,
which are usually applied to systems possessing discretized energy states,
ignore or simplify such tunneling contributions, often through use of the rotating wave approximation (RWA).
The main problem created by the RWA is that
a system treated under this approximation will not satisfy the fluctuation-dissipation theorem,
and thus it may introduce significant error in the evolution of the system toward equilibrium.
In the classical limit, with $\hbar$ tending to zero, $C(t)$ is always positive.

We assume that spectral density $J(\omega)$ has an Ohmic form
with a Lorentzian cutoff, i.e.,
\begin{equation}
 J(\omega) = \frac{m\zeta}{\pi}\frac{\gamma^2\omega}{\gamma^2+\omega^2},
\label{Jform}
\end{equation}
where $\zeta$ is the system-bath coupling strength, which represents the magnitude of damping, and
$\gamma$ is the width of the spectral density of the bath mode.
The canonical and symmetrized correlation functions then become
\begin{equation}
\Psi(t) = m \zeta \gamma{\rm e}^{ - \gamma \left| t \right|},
\label{eq:Psi}
\end{equation}
and
\begin{align}
 C(t)=\frac{m\hbar\zeta\gamma^2}{2}\left[ \cot\left(\frac{\bhr}{2}\right) e^{-\gamma|t|}
	- \frac{1}{\beta\hbar}\sum_{k=1} \frac{4\nu_k}{\gamma^2-\nu_k^2}e^{-\nu_k|t|} \right],
\label{eq:C}
\end{align}
where $\nu_k=2\pi k/\beta\hbar$ is the $k$-th Matsubara frequency.\cite{Tanimura90}
Note that in the high temperature limit, $\beta\hbar\gamma \ll 1$,
the noise correlation function reduces to $C(t)\approx m\zeta \gamma e^{-\gamma|t|}/\beta$.
This indicates that the heat bath oscillators interact with the system in the form of Gaussian-Markovian noise. 

The reduced hierarchy equations of motion (HEOM) can be obtained
by considering the time derivative of the reduced density matrix elements
given in eq (\ref{rho_red}) in the Wigner representation,\cite{TJPSJ2006}
as a generalization of the Gaussian-Markovian quantum Fokker-Planck equation
introduced by Tanimura and Wolynes\cite{TanimuraWolynes1,TanimuraWolynes2}.
If we choose $K$ so as to satisfy $\nu_K=2\pi K/(\beta\hbar) \gg \omega_c$,
where $\omega_c$ is the characteristic frequency of the system,
the factor ${\rm e}^{-\nu_k |t|}$ in eq (\ref{eq:C}) can be replaced with the Dirac delta function,
using the approximation $\nu_k{\rm e}^{-\nu_k |t|}/2\simeq \delta( t ) \quad ({\rm for} \ \ k \geq  K+1)$, with negligible error.
For the kernel equations (\ref{eq:Psi}) and (\ref{eq:C}),
the number of hierarchy elements for $\gamma$ is denoted by $n$,
and the number of $k$th Matsubara frequencies is denoted by $j_k$.\cite{TJPSJ2006,ITJSPS2005}
Then the HEOM can be expressed as\cite{STJPCA2011,STJPSJ2013,STNJP2013}
\begin{align}
 \frac{\partial}{\partial t}{W}_\jk^{(n)}(q,p;t) &= -\left[ \mathcal{L}_{QM} + \hat{\Xi} + n\gamma + \sum_{k=1}^K j_k\nu_k \right] W_\jk^{(n)}(q,p;t) \nonumber \\
	& + \hat{\Phi} \left[ W_\jk^{(n+1)}(q,p;t) + \sum_{k=1}^K W_\jkp^{(n)}(q,p;t) \right] \nonumber \\
	& + n\gamma \hat{\Theta}_0 W_\jk^{(n-1)}(q,p;t) \nonumber \\
	& + \sum_{k=1}^K j_k\nu_k \hat{\Theta}_k W_\jkm^{(n)}(q,p;t), 
\label{heom_wig}
\end{align}
where the quantum Liouvillian in the Wigner representation is expressed as \cite{Frensley}
\begin{equation}
 -\mathcal{L}_{QM}W(q,p) = -\frac{p}{m}\ddq W(q,p) - \int_{-\infty}^\infty \frac{dp'}{2\pi\hbar^2}U_W(q,p-p';t)W(q,p'), 
\end{equation}
with $U_W(q,p;t) = 2\int_0^\infty dr \sin(pr/\hbar)[ U(q+r/2;t)-U(q-r/2;t)]$.
The other operators are defined as
\begin{equation}
 \hat{\Phi} \equiv \ddp,
\end{equation}
\begin{equation}
 \hat{\Theta}_0 \equiv \zeta \left[ p + \frac{\mhr}{2}\cot \left( \frac{\bhr}{2}\right) \ddp \right],
\end{equation}
\begin{equation}
 \hat{\Theta}_k \equiv \frac{2m\gamma^2\zeta}{\beta (\nu_k^2-\gamma^2)}\ddp,
\end{equation}
for $k>0$, and
\begin{equation}
 \hat{\Xi} \equiv -\frac{m\zeta}{\beta}\left[ 1 - \frac{\bhr}{2}\cot\left( \frac{\bhr}{2}\right) - \sum_{k=1}^K \frac{2\gamma^2}{\nu_k^2-\gamma^2}\right]\ddpp.
\end{equation}
Note that the 0th element is identical to the Wigner function $W_{0,0,\cdots,0}^{(0)}(q,p;t)\equiv W(q,p;t)$,
and the other elements are introduced in the numerical calculations
in order to treat the non-perturbative, non-Markovian system-bath interaction.
Although these elements do not have direct physical meaning,
they allow us to take into account the quantum coherence and entanglement between the system and the bath.\cite{Dijkstra10, Nori12}
The importance of the system-bath coherence was pointed out in the context of correlated initial conditions,\cite{Grabert}
and it was shown with a nonlinear response theory approach that the system-bath coherence plays an essential role
in the case that the system is driven by a time-dependent external force.\cite{TJPSJ2006}

The HEOM consist of an infinite number of equations,
but it can be truncated at finite order with negligible error.\cite{TanimuraWolynes1,TanimuraWolynes2}
Essentially, the condition necessary for the error introduced by the truncation to be negligibly small is that
the total number of hierarchy elements or the total number of Matsubara frequencies retained be sufficiently large.
Explicitly, it can be shown that the condition $N\equiv n+\sum_{k=1}^K j_k \gg \omega_c/\min (\gamma ,\nu_1)$ is sufficient for this purpose.
\cite{ITJSPS2005} For the Caldeira-Leggett Hamiltonian, the equations of motion are then truncated by using the "terminator"s,
expressed in the Wigner representation as\cite{STJPCA2011,STJPSJ2013,STNJP2013}
\begin{equation}
 \frac{\partial}{\partial t}{W}_\jk^{(n)}(q,p;t) = -\left( \mathcal{L}_{QM} + \hat{\Xi} \right) W_\jk^{(n)}(q,p;t). 
\label{wig_term}
\end{equation}
with $n+\sum_{k=1}^K j_k = N$.
The HEOM formalism can be used to treat a strong system-bath coupling non-perturbatively. 
It is ideal for studying quantum transport systems when employing the Wigner representation,
because it allows the treatment of continuous systems utilizing open boundary conditions and periodic boundary conditions\cite{Frensley}.
In addition, the formalism can accommodate the inclusion of an arbitrary time-dependent external field
while still accounting for the system-bath coherence.
Note that such coherence cannot be described if we assume that the state of the total system takes a factorized system-bath form.\cite{TJPSJ2006}
Such system-bath coupling features are necessary to properly treat quantum ratchet systems. 
In the white noise (or Markovian) limit, $\gamma \rightarrow \infty$,
which is taken after imposing the high temperature approximation, $\beta \hbar \gamma \ll 1$,
the quantum Fokker-Planck equation can be derived in a similar form as the Kramers equation,\cite{CLPhysica1983,Waxman1985}
which is identical to the quantum master equation without RWA.\cite{Carmichael1999}
Because we assume $\beta \hbar \gamma \ll 1$ with $\gamma \to \infty$,
this equation cannot be applied to low-temperature systems, where quantum effects play a major role. 

The Wigner distribution function reduces to the classical one in the limit $\hbar \to 0$,
and hence we can directly compare the quantum results with the classical results.
The classical HEOM, which is derived from eq (\ref{heom_wig})
by setting $\hbar=0$, is given by\cite{TanimuraWolynes1,TanimuraWolynes2}
\begin{align}
 \frac{\partial}{\partial t}{W}^{(n)}(q,p;t)  =& -\left( \mathcal{L}_{CL} + n\gamma \right) W^{(n)}(q,p;t) \nonumber \\ 
	& + \ddp W^{(n+1)}(q,p;t) \nonumber \\
	&+ n\gamma\zeta\left( p + \frac{m}{\beta}\ddp \right) W^{(n-1)}(q,p;t), 
\label{heom_cl}
\end{align}
and the classical terminator is
\begin{align}
  \frac{\partial}{\partial t}{W}^{(N)}(q,p;t) =& -\left( \mathcal{L}_{CL} + N\gamma \right) W^{(N)}(q,p;t) \nonumber \\ 
	& + \zeta\ddp \left( p + \frac{m}{\beta}\ddp \right)W^{(N)}(q,p;t) \nonumber \\
	& + N\gamma\zeta\left( p + \frac{m}{\beta}\ddp \right) W^{(N-1)}(q,p;t).
\label{cl_term}
\end{align}
The classical Liouvillian is defined by
\begin{equation}
 -\mathcal{L}_{CL}W(q,p) = -\frac{p}{m}\ddq W(q,p) + \frac{\partial U(q;t)}{\partial q}\ddp W(q,p).
\label{L_cl}
\end{equation}
The classical equation of motion is helpful,
because knowing the classical limit allows us to identify the purely quantum mechanical effects \cite{TanimuraWolynes2,STJPCA2011}.
Note that the above equations of motion can also be derived from the classical Langevin equation,
where the classical fluctuation dissipation theorem is hold.\cite{TanimuraWolynes1}
Then, in the white noise (or Markovian) limit, the above equations reduce to the Kramers equation.

In Sec. 4, we use the eqs  (\ref{heom_wig})-(\ref{wig_term}) and eqs  (\ref{heom_cl})-(\ref{L_cl})
to calculate quantum and classical ratchet currents, respectively.
Because we are dealing with the distribution function for both quantum and classical cases,
there is no need to sample kinetic trajectories,
which is necessary in approaches employing the Langevin equation and in some kinetic approaches.
This is convenient if we wish to calculate the current at low temperature, even in the classical case,
because this sampling becomes inefficient due to the localization of the particle trajectories trapped in the potential.
Another important aspect of the present methodology is that
it allows elucidation of the dynamical behavior of the system through the time evolution of the distribution function in the phase space.
In addition to the dependence of the current on the temperature, the barrier height and the system-bath coupling,
we analyze the mechanism of ratchet rectification using the Wigner and classical distribution functions. 

\begin{figure}
\includegraphics[width=10cm]{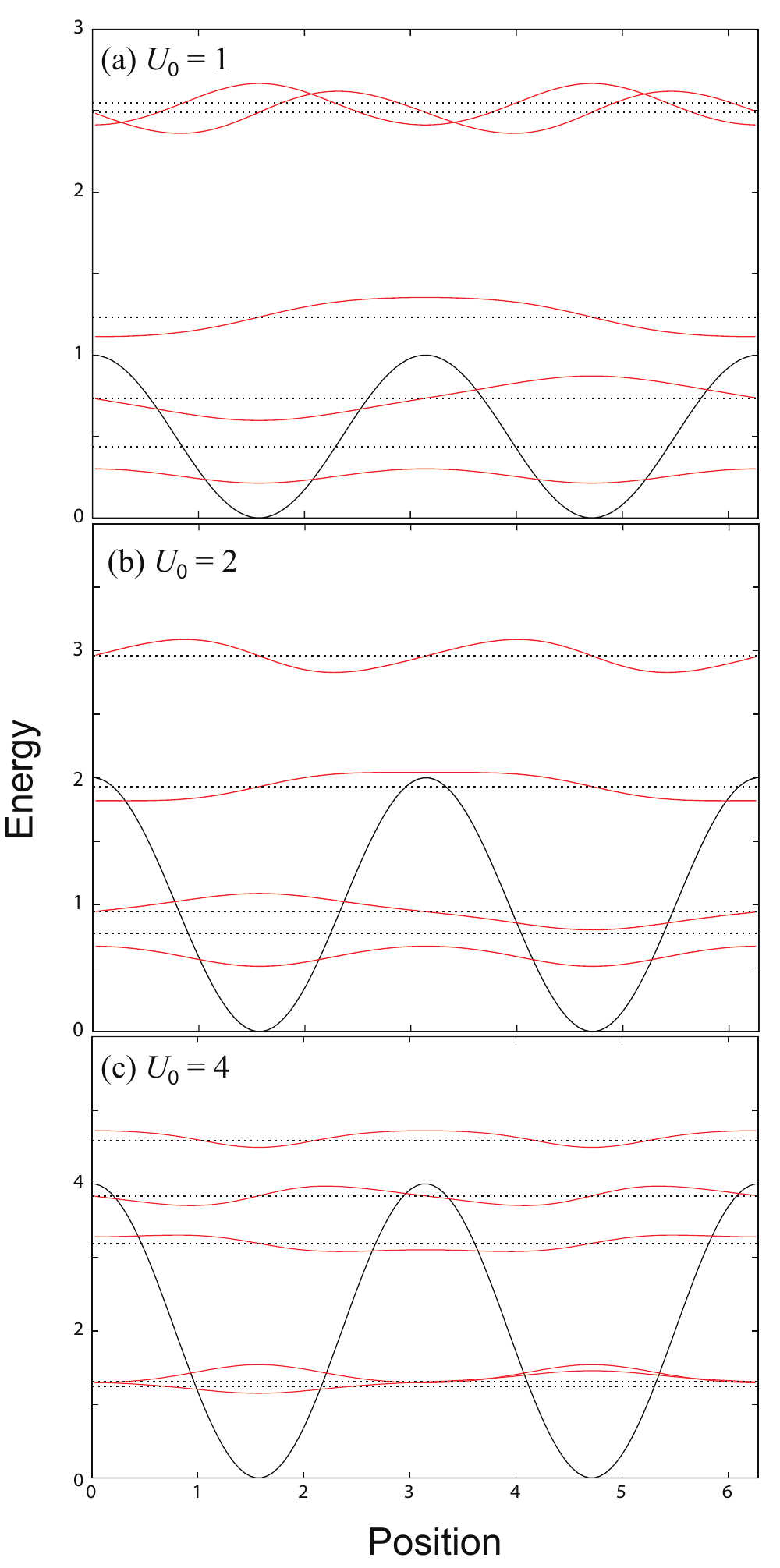}
\caption{The $U_0\cos^2({\bar q})$ potential for the cases of (a) a low barrier, $U_0=1$, (b) an intermediate barrier, $U_0=2$, and (c) a high barrier, $U_0=4$.
In each case, eigenenergies and eigenfunctions with periodic boundary conditions are plotted as the black dashed lines and the red curves, respectively.}
\label{Potential}
\end{figure}

\section{Potential, Force and Symmetry}
In the present study, we consider the one-dimensional symmetric potential (see Figure \ref{Potential})
\begin{equation}
  U(q) = U_0 \cos^2(\kappa q)
\end{equation}
and the biharmonic force
\begin{equation}
  F(t) = F_1\cos (\Omega t) + F_2\cos (2\Omega t + \theta ),
\label{F_t}
\end{equation}
where $U_0$ is the barrier height, $\kappa$ is the wave number,
and $F_1$, $F_2$, $\Omega$ and $\theta$ denote the amplitudes, frequency, and the phase difference of the biharmonic forces, respectively. \cite{Seeger66,Breymayer81,Breymayer82,Breymayer84,Marchesoni86,Hanggi98EPL,FlachPRA07,Hanggi09EPL,Grossmann11, Hanggi09AJP,Machura10PRE}
Experimentally, such a situation has been realized for fluxons in long Josephson junctions\cite{Ustinov04} and
for cold rubidium and cesium atoms in optical lattices.\cite{Shiavoni03,Renzoni05,Salger09}

Before reporting the results of our calculations of the ratchet current,
it is important to mention two symmetry conditions under which the current vanishes.\cite{FlachPRL00,FlachERL01,ZolotaryukPRE02,ReimannPRL01,Quintero10}
The first one is the time-shift symmetry defined by the transformation $(q,p,t)\to (-q,-p,t+T/2)$, 
where $T$ is the period of the external force,
and the second one is the time-reversal symmetry defined by the transformation $(q,p,t)\to (q,-p,-t)$. 
When the equation of motion of the system is invariant under either of these two symmetry transformations,
the current vanishes, because each of these transformations reverses the sign of the momentum.
For the external force given by eq (\ref{F_t}) 
and for the quantum and classical HEOM appearing in eqs (\ref{heom_wig})-(\ref{wig_term}) and eqs (\ref{heom_cl})-(\ref{L_cl}), respectively,
time-shift symmetry exists if $F_1=0$ or $F_2=0$.
Time-reversal symmetry is broken if $\theta \ne 0$ or $\pi$ in the Hamiltonian case, i.e. if $\zeta=0$,
while, in the presence of dissipation, this symmetry is always broken, which implies that directed motion can exist for arbitrary $\theta$.
For the classical overdamped Langevin equation with white noise,
the system possesses time-reversal symmetry if $\theta=\pi/2$ or $3\pi/2$.
The overdamped (Smoluchowski) case can be studied within the HEOM formalism
by choosing a large value for $\zeta$ with $\gamma\gg \Omega$ in the classical case
and by choosing a large value of $\zeta$ with $\gamma\gg \Omega$ and $\beta \hbar \gamma \ll 1$ (i.e. the high temperature limit) in the quantum case. In such cases, we expect the calculated current to vanish if $\theta=\pi/2$ or $3\pi/2$.

\section{Numerical Results}
We numerically integrated the HEOM in the form of finite difference equations
using the fourth-order Runge-Kutta method.
We also imposed the periodical boundary condition $W(p,q)=W(p,q+2\pi/\kappa)$. 
The spatial derivative of the kinetic term in the Liouville operator, $-(p/m) \partial W(p,q)/\partial q$,
was approximated using a third-order left-handed or right-handed difference scheme,
depending on the sign of the momentum,
while other derivatives with respect to $p$ were approximated using a third-order center difference scheme.\cite{STNJP2013} 
We employed the following units of the length, momentum, and frequency: $q_r= 1/\kappa$, $p_r= \hbar \kappa$ and $\omega_r= \hbar \kappa^2/m$.
Note that with these units, we have $m = \hbar = \kappa = 1$.
The dimensionless position and momentum then are given by ${\bar q}=q/q_r$ and ${\bar p}=p/p_r$.
The mesh sizes for the position and the momentum were chosen in the ranges $0.0524< \Delta {\bar q}< 0.1047$ and $0.04< \Delta {\bar p}< 0.10$.
The depth of the hierarchy and the number of Matsubara frequencies were chosen so as to satisfy $N \in \{ 5-10\}$ and $K\in \{ 1-2\}$.
In this study, we fixed the inverse of the noise correlation time to $\gamma = 1.0\omega_r$ 
and the amplitude and frequency of the biharmonic force to $F_1= F_2= 0.20\hbar\omega_r/q_r$ and $\Omega = 1.0\omega_r$, respectively.
Three values were used for the height of potential barrier, $U_0= \hbar \omega_r, 2\hbar \omega_r$, and 4$\hbar \omega_r$ (see Figure \ref{Potential}).

Employing the biharmonic force given in eq (\ref{F_t}),
we integrated the classical and quantum HEOM until the distributions reached the steady state.
Then, the classical and quantum directed currents (or the ratchet current) were evaluated from the distribution function as
\begin{equation}
 J\equiv \lim_{t\to \infty} \frac{1}{T}\int_t^{t+T}dt' \langle p(t')\rangle ,
\end{equation}
where
\begin{equation}
 \langle p(t) \rangle = \int dqdp~pW(q,p;t).
\end{equation}
We repeated the calculation for fixed physical conditions,
while varying $\theta$ in the range $0 < \theta < 2\pi$.
From these calculations, we find that as a function of $\theta$, the current roughly takes the following form:\cite{Shiavoni03,Hanggi09AJP}
\begin{equation}
J(\theta) = J_{max} \sin (\theta-\theta_{0}),
\label{jmax}
\end{equation} 
where $J_{max}$ and $\theta_{0}$ are the maximum value of the current and the phase lag evaluated from the numerical simulations, respectively.
Below we investigate the dependence of the current on the temperature, the barrier height and the bath coupling strength
through the dependences of $J_{max}$ and $\theta_{0}$ on these quantities.
To gain further insight into the mechanism of ratchet rectification,
we also present plots of the Wigner and classical distribution functions where helpful.

\begin{figure}
\includegraphics[width=15cm]{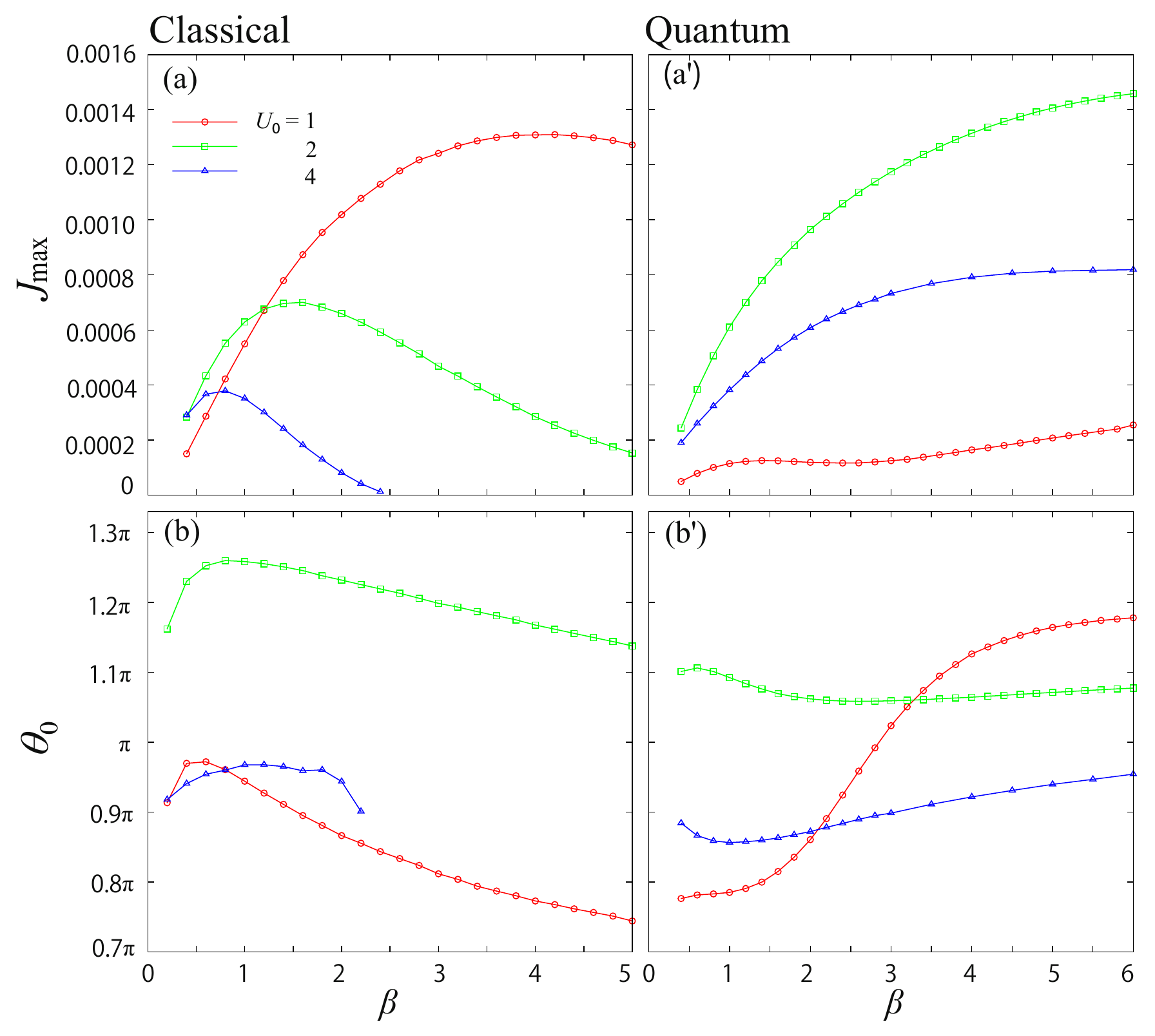}
\caption{The maximum current in (a) the classical and (a') the quantum case and the corresponding phase lag
in (b) the classical and (b') the quantum case
as functions of the inverse temperature, $\beta$, for three values of the barrier height $U_0$.
The coupling strength here is $\zeta= 0.10\omega_r$.}
\label{Max_current}
\end{figure}

\begin{figure}
\includegraphics[width=13cm]{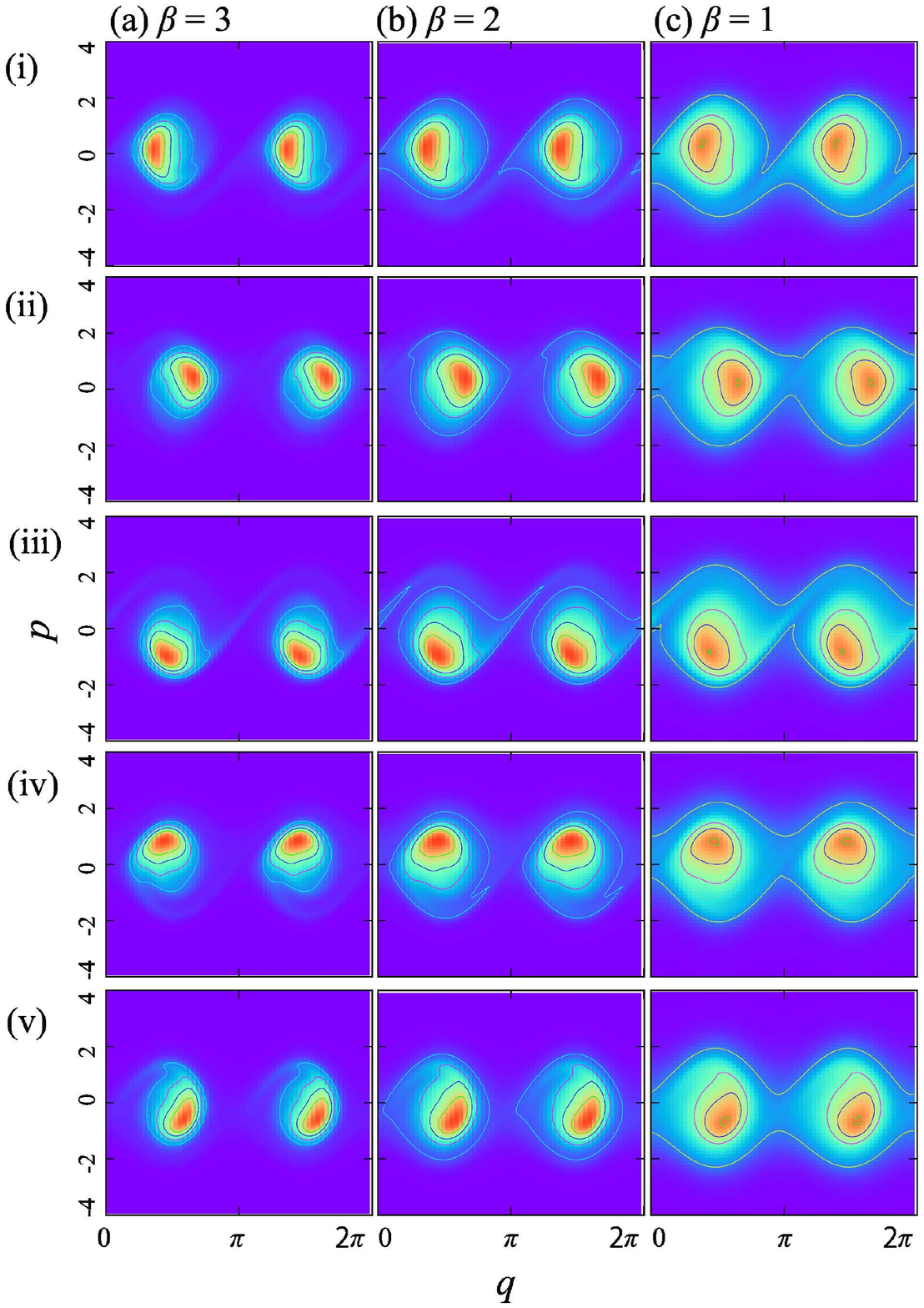}
\caption{Snapshots of the classical distribution for $\theta = 0.7\pi$ in the case of intermediate barrier height, $U_0=2$,
for three values of the inverse temperature: (a) low temperature, $\beta=3$; (b) intermediate temperature, $\beta=2$; (c) high temperature, $\beta=1$.
The snapshots correspond to the following times: $t=$ (i) $0$, (ii) $0.4\pi$, (iii) $0.8\pi$, (iv) $1.2\pi$, (v) $1.6\pi$ ($1/\Omega$).}
	\label{Classical_ditribV2}
\end{figure}

\subsection{Temperature Effects: the Classical Case}
Figure \ref{Max_current} displays (a) the maximum value, $J_{max}$, and (b) the phase lag, $\theta_{0}$,
of the current as functions of the inverse temperature, $\beta$, for three values of the barrier height, $U_0$,
evaluated using the classical HEOM given in eqs (\ref{heom_cl})-(\ref{L_cl}).
Here, we chose the weak coupling strength $\zeta= 0.10\omega_r$.
In all three cases depicted in Figure \ref{Max_current}(a),
it is seen that the maximum current realizes a maximum value at some intermediate value of the inverse temperature.
The value of $\beta$ at which this maximum is realized is found to increase as $U_{0}$ decreases.
As shown in Figure \ref{Max_current}(b), similar profile is found for the phase lag $\theta_{0}$ as a function of $\beta$, although the peak positions are different.

The behavior depicted in Figure \ref{Max_current} can be understood in terms of the thermal activation of a particle.
To illustrate this, in Figure \ref{Classical_ditribV2},
we plot snapshots of the classical distribution function for (a) low, (b) intermediate, and (c) high temperature cases
at five different values of the time for one cycle. 
In Figures \ref{Classical_ditribV2}(a-i) and (b-i),
it is seen that when the temperature takes a low or intermediate value,
the distribution function is concentrated in the potential wells.
The distributions are periodically accelerated by the biharmonic force and rotate clockwise around the bottom of each potential well. 
Figures \ref{Classical_ditribV2}(a-ii) and (b-ii) reveal that when the distribution approaches the left side of the barrier,
a small part of the distribution with positive momentum is transferred to the right potential well by crossing the barrier. 
This can be seen as the elongation of the distribution on the right side. 
In Figures \ref{Classical_ditribV2}(a-iii) and (b-iii), it is seen that
while the distribution begins to flow in the positive direction,
the sign of the biharmonic force changes, and the main part of the distribution moves away from the barrier.
The part of the distribution that crosses the barrier flows into the right potential and rotates clockwise in a spiral form while losing energy, due to dissipation. 
Figures \ref{Classical_ditribV2}(a-iv) and (b-iv) show the flow stops shortly after the biharmonic force change direction, with a small time delay.
In Figures \ref{Classical_ditribV2}(a-v) and (b-v), it is seen that when the distribution approaches the left-hand side of the barrier,
a part of the distribution with negative momentum also flows in the opposite direction, 
but the effect of the acceleration caused by the biharmonic force is smaller in this case than in the positive case,
due to the asymmetric influence of the biharmonic force on the system dynamics with the phase lag $\theta_0$.

In the intermediate temperature case depicted in Figure \ref{Classical_ditribV2}(b),
the distribution is broadened compared with Figure \ref{Classical_ditribV2}(a) due to thermal effects.
The current is greater than in the low temperature case,
because the weight of the distribution at energies above the barrier under the biharmonic force is greater.
The response of the distribution to the external force is also delayed due to the broadening of the distribution in the $p$ direction.
Because the phase lag, $\theta_0$, changes in accordance with the timing of the excitation created by $F_1\cos (\Omega t)q$ and $F_2\cos (2\Omega t + \theta - \theta_0 )q$ through the gradients of the potential, this delay may be the reason that $\theta_0$ increases as $\beta$ decreases up to the maximum point.

In the high temperature case considered in Figure \ref{Classical_ditribV2} (c), 
the distribution is populated even at the top of barrier. 
This dispersion of the distribution indicates that the effect of the ratchet
potential is suppressed at high temperatures. This causes the spatio-temporal
asymmetry, which is necessary for ratchet current, to be weak, and as a
result, the current decreases as the temperature increases.
Since the activation energy increases as the barrier increases,
the maxima of the peaks shift toward higher temperature as $U_0$ increases.
In the high temperature case, because the distribution is dispersed over the potential,
the distribution can flow to the neighboring potentials
both in the positive and negative directions with only a small delay.
This can be understood from Figure \ref{Classical_ditribV2}(c-ii),
where it is seen that the current flows quickly following the movement of the distribution.
Thus, $\theta_0$ decreases as the temperature increases,
although the peak positions are different from those of $J_{max}$.
This is due to the fact that there is only an indirect relation between the current and the phase. 

\subsection{Temperature Effects: the Quantum Case}
Next, we consider the quantum mechanical case under the same physical conditions as in the classical case considered above.
Figure \ref{Max_current} displays (a') the maximum value, $J_{max}$,
and (b') the phase lags, $\theta_{0}$, of the current as functions of the inverse temperature, $\beta$,
for three values of the barrier height, $U_0$,
calculated using the quantum mechanical HEOM given in eqs  (\ref{heom_wig})-(\ref{wig_term}). 
It is seen that the profiles of $J_{max}$ and $\theta_{0}$
differ significantly from those in the classical case.
In the intermediate and high barrier cases, 
The figure also shows that the quantum mechanical $J_{max}$ is significantly larger than the classical one
for the cases of an intermediate and high barrier in the low temperature regime ($\beta > 2$),
while they exhibit similar behavior in the high temperature regime ($\beta < 1$).

To see a role of quantum effects, we consider the crossover temperature
at which the classical thermal-activated regime crosses over to the quantum tunneling regime
with regard to chemical reaction rates\cite{HGIW1985}
for the bath spectral density, $J(\omega)$, given in eq (\ref{Jform})
with the potential heights $U_{0}= 1, 2,$ and $4$.
The evaluated crossover temperatures are at $\beta_c$=4.5, 3.2, and 2.2, respectively.
These values of $\beta_{c}$ are much smaller than those at which we observed the difference between the quantum and classical cases.
This suggests that our analysis based on the reaction rate may not apply directory,
because here we considered $J_{max}$ under the external driving force while altering the phase.

A distinctive feature of the quantum mechanical results is found
in the low barrier case (the red curve) in Figure \ref{Max_current}(a').
One may expect that the ratchet current in the quantum mechanical case
is larger than that in the classical case, especially at low temperatures due to tunneling.
However, our results in the low barrier case presented in Figure \ref{Max_current}(a')
reveal that in fact this is not the case.

\begin{figure}
	\includegraphics[width=13cm]{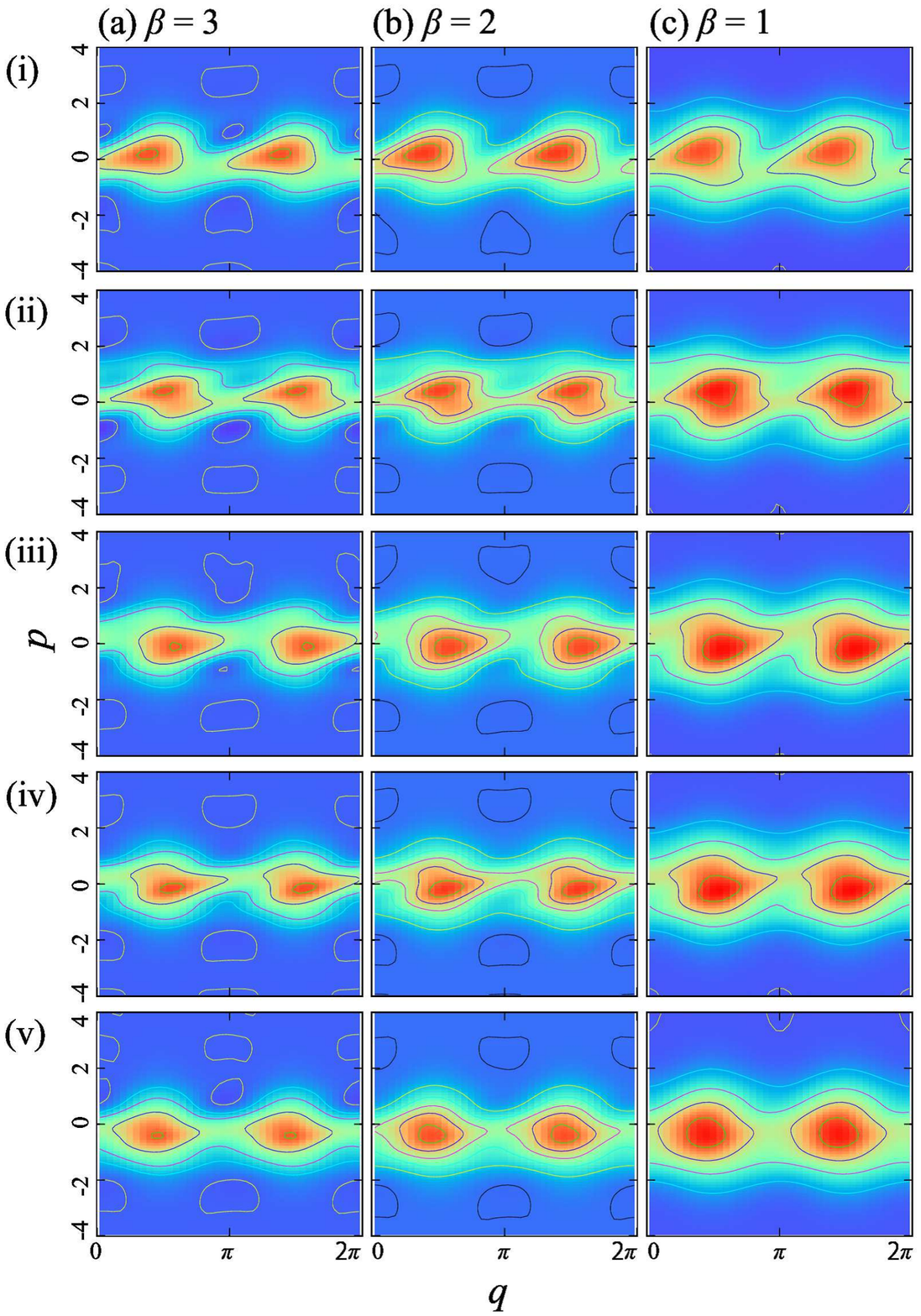}
	\caption{Snapshots of the quantum distribution for $\theta = 0.4\pi$ in the case of low barrier height, $U_0=1$, calculated using the HEOM
for three values of the inverse temperature: (a) low temperature, $\beta=3$; (b) intermediate temperature, $\beta=2$; (c) high temperature, $\beta=1$.
These snapshots correspond to the following times: $t=$ (i) $0$, (ii) $0.4\pi$, (iii) $0.8\pi$, (iv) $1.2\pi$, (v) $1.6\pi$ ($1/\Omega$).}
	\label{quantum_ditribU1}
\end{figure}

\begin{figure}
	\includegraphics[width=13cm]{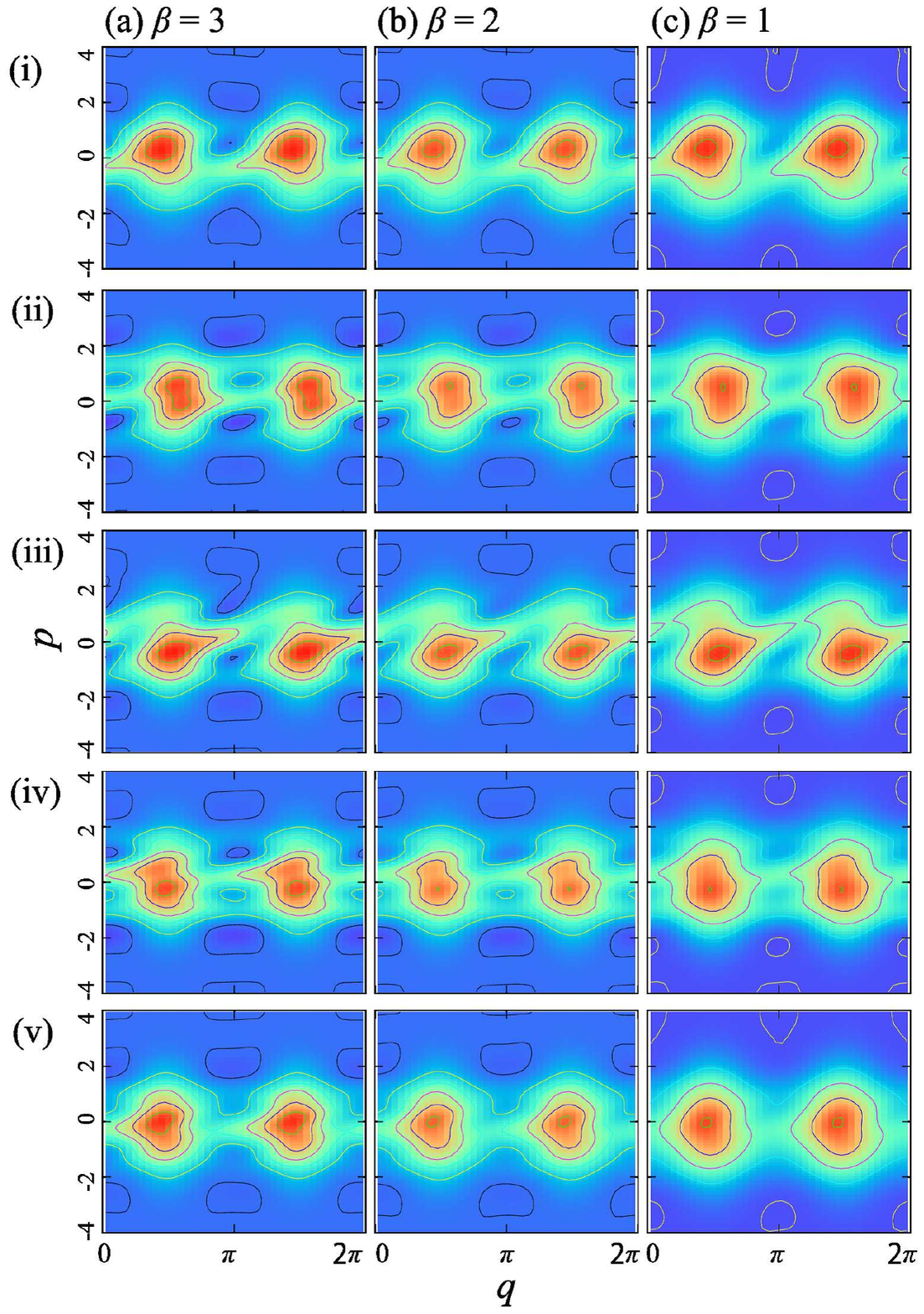}
	\caption{Snapshots of the quantum distribution for $\theta = 0.6\pi$ in the case of intermediate barrier height, $U_0=2$,
for the following values of $\beta$: (a) low temperature, $\beta=3$; (b) intermediate temperature, $\beta=2$; (c) high temperature, $\beta=1$.
These snapshots correspond to the following times: $t=$ (i) $0$, (ii) $0.4\pi$, (iii) $0.8\pi$, (iv) $1.2\pi$, (v) $1.6\pi$ ($1/\Omega$).}
	\label{quantum_ditribU2}
\end{figure}

To elucidate the reason for the behavior depicted in Figure \ref{Max_current}(a') for the case $U_{0}= 1$,
we consider the Wigner distribution in the low and intermediate barrier cases
for three values of the temperature in Figures \ref{quantum_ditribU1} and \ref{quantum_ditribU2}, respectively.
In both cases, the Wigner distribution function is dispersed over the potential,
and it is negative in some regions (the regions inside the white and black loops) at low temperature, ($\beta \ge 2$).
The Wigner function for the quantum system is not positive definite,
and in cases that the tunneling process is important,
the region in which it takes negative values is larger.\cite{Wigner84}
This indicates that delocalization of the distribution in the low barrier case
depicted in Fig \ref{quantum_ditribU1} (a) arises from tunneling.
This can be understood from Figure \ref{Potential}(a),
where the ground state in the low potential case is seen to have a large population under the barrier. 
The tunneling contribution induced by biharmonic perturbation may be suppressed at high temperature,
because the temperature enters the theory through $\beta \hbar$,
and hence the classical limit, $\hbar \to 0$, is equivalent to the high temperature limit, $\beta \to 0$.
In this case, however,
thermal activation causing flow over the potential barrier is significant.
For this reason, the distribution is dispersed for all temperatures. 
As shown in the high temperature classical case depicted in Figure \ref{Classical_ditribV2}(b),
the effect of the ratchet potential is weak when the distribution function is dispersed.
This is the reason that the ratchet current is smaller in the quantum case than the classical case if the barrier is low. 

In the cases of intermediate and high barriers, depicted in Figure \ref{Max_current}, tunneling also plays a significant role.
As shown in Figure \ref{quantum_ditribU2}, the Wigner distribution is not dispersed when the barrier is high.
As a result, the ratchet mechanism is effective with the help of tunneling,
and thus the quantum mechanical current is larger than the classical current.
However, when the temperature becomes very high,
the quantum coherence is lost due to the thermal noise,
and as a result, the quantum value approaches the classical one. 

As shown in Figure \ref{Max_current}(b'), the phase lag has a critical point in the high temperature region ($\beta < 1.5$)
as in the classical case depicted in Figure \ref{Max_current}(b).
However, in the low temperature region($\beta > 2.5$),
the phase is seen to increase monotonically in the quantum mechanical case,
while it decreases monotonically in the classical case.
We carefully investigated the dynamics of the Wigner distribution in the low temperature region
by studying snapshots of it and the corresponding current using fine time slices.
We thus found that the transfer of the distribution through tunneling takes a longer time at low temperature.
This is because thermal transitions between the ground and excited states,
which is the cause of tunneling transition induced by biharmonic perturbation, may be suppressed at low temperature.
Because the net current flows slowly in the low temperature case,
the phase lag that controls the timing of the biharmonic force increases when the temperature decreases.
For higher barriers, the tunneling plays a minor role, and thus the increase of the phase is smaller than in the low barrier case.

Note that, as discussed in Ref.\cite{HanggiPRB89},
quantum diffusion may be suppressed by quantum reflection.
In the cases we studied, however,
the curvature at the top of the barrier and in the potential well are equal,
and the force that creates the difference between the curvatures is small.\cite{PollakJCP03}
Thus, in the present case, quantum reflection is not the cause of the suppression. 

\begin{figure}
\includegraphics[width=12cm]{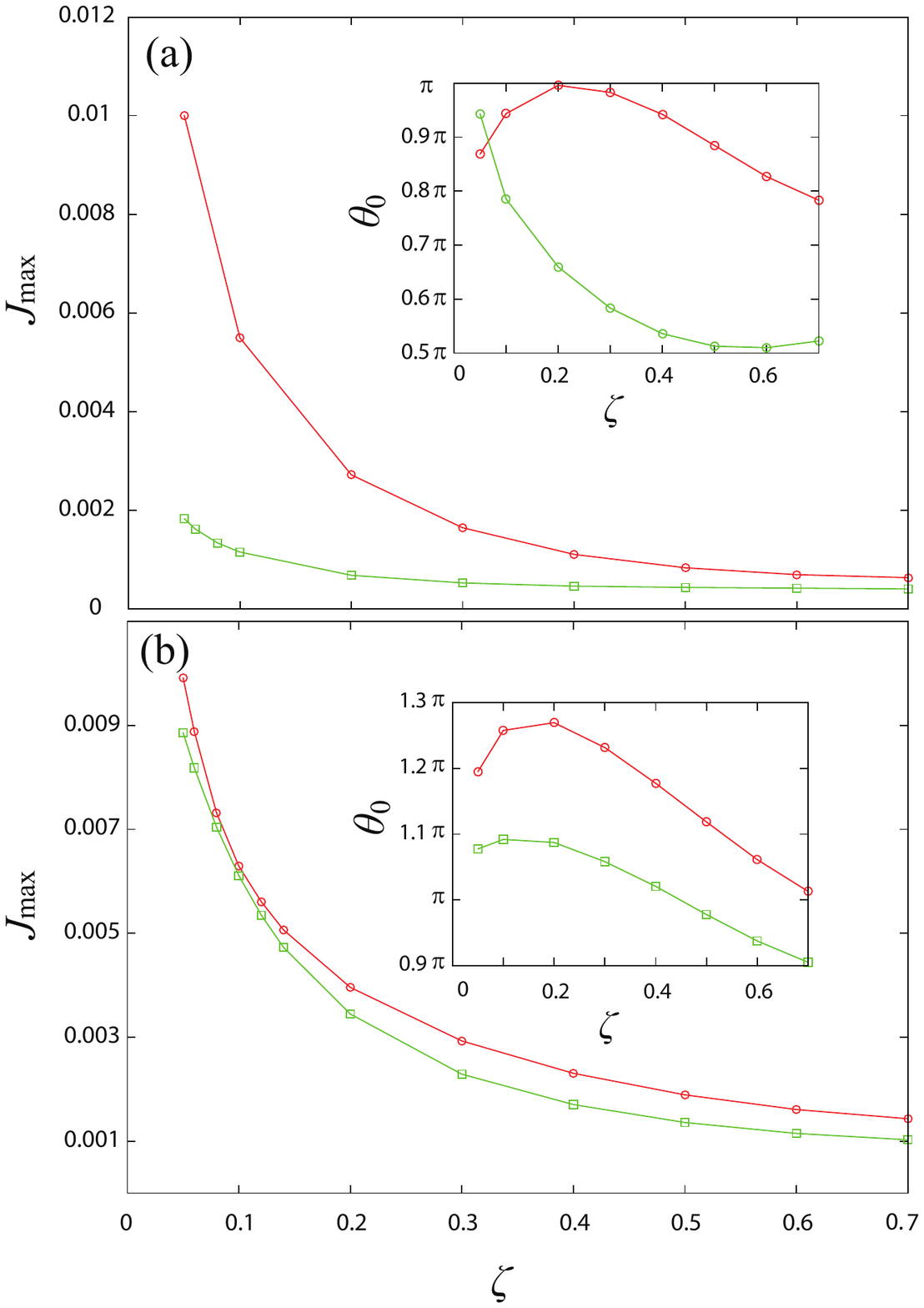}
\caption{The maximum ratchet current in the classical (red curve) and quantum mechanical (green curve) cases
for (a) a low barrier, $U_0=1$, and (b) an intermediate barrier , $U_0=2$, with $\beta=1$. The insets depict the phase lag $\theta_{0}$ as a function of $\zeta$.}
\label{zeta_current}
\end{figure}

\begin{figure}
\includegraphics[width=17cm]{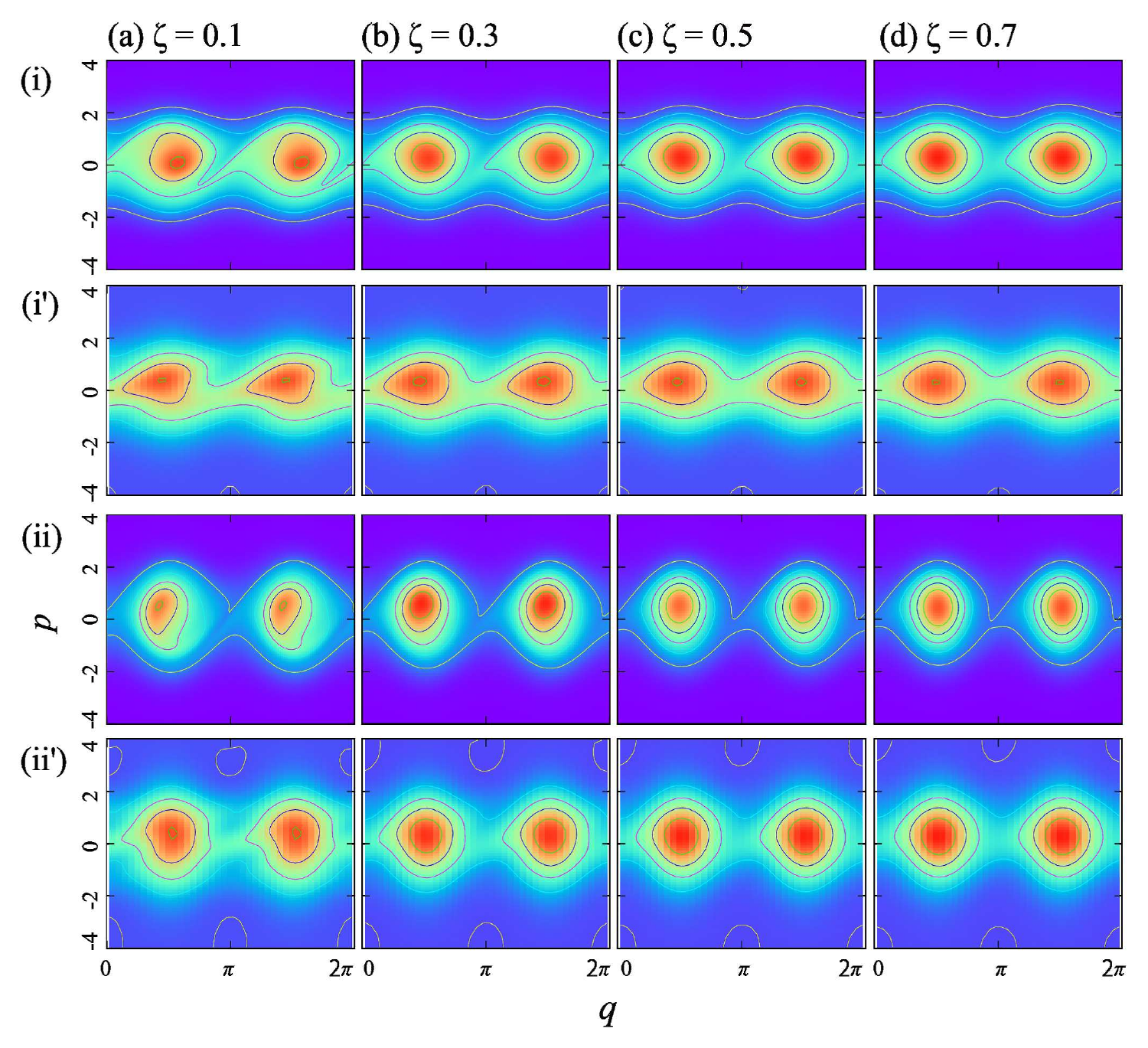}
\caption{Snapshots of the classical (upper panel) and quantum mechanical (lower panel) distribution functions
at the fixed time $t=2\pi$ ($1/\Omega$) for four values of the coupling strength: $\zeta=$ (a) 0.1, (b) 0.3, (c) 0.5, (d) 0.7.
The upper panels, (i) and (ii), describe the classical cases,
whereas the lower panels, (i') and (ii'), describe the quantum cases for the case of a low barrier, $U_0=1$, and an intermediate barrier, $U_0=2$, respectively. }
	\label{zeta_ditrib}
\end{figure}

\subsection{Varying the Coupling Strength}
We next investigate the change of the ratchet current as the system-bath coupling strength $\zeta$ is varied
by taking advantage of the fact that the HEOM constitute a non-perturbative theory.
In Figure \ref{zeta_current}, we plot the classical (red curve) and quantum (green curve) ratchet current
for (a) the low barrier case, $U_0=1$, and (b) the intermediate barrier case, $U_0=2$, with $\beta=1$.
In both cases, the classical and quantum currents seem to approach the same value for large $\zeta$,
although the quantum values reach the large $\zeta$ limit more rapidly.
To compare the two cases, we plot the quantum and classical distribution functions in Figure \ref{zeta_ditrib}.
It is seen that in the low barrier case, the quantum distribution is dispersed, as in the case depicted in Figure \ref{quantum_ditribU1}.
Moreover, in the quantum mechanical case, the distribution in the $p$ direction becomes a symmetric Gaussian for $\zeta>0.3$,
while there is some deviation from Gaussian in the classical case, due to the external perturbation.
This Gaussian feature arises from quantum tunneling,
as can be observed more clearly in the low barrier case.
It is interesting that the situation in which there exists a symmetric Gaussian distribution in momentum space
is similar to that described by the classical Smoluchowski equation,
which can be derived in the overdamped limit from the Kramers equation.
In the overdamped limit, the distribution of the momentum direction is Gaussian.
In the presently considered case, we find that such a situation arises due to the effect of tunneling
even if the system-bath coupling is weak.
Therefore, it is natural to call this situation a tunneling-induced Smoluchowski limit.

The phase approaches $\pi/2$ as $\zeta$ increases, as depicted in the inset of Figure \ref{zeta_current}(ii).
This is similar to the situation for the classical Smoluchowski equation. 
For $U_0=2$, the quantum and classical cases exhibit similar $\zeta$ dependence.
This is due to the fact that the dispersion of the distribution through tunneling does not occur in the higher barrier case,
as depicted in Figure \ref{quantum_ditribU2}.
The quantum values approach the classical values for large $\zeta$,
because in that case, the quantum coherence is destroyed by the system-bath coupling.

\section{Concluding Remarks}
The classical and quantum mechanical ratchet currents in a system driven by a biharmonic force
were calculated in a rigorous manner for the first time using the reduced hierarchy equations of motion (HEOM)
in the Wigner representation over a wide range of values of the temperature, the system-bath coupling strength, and the potential height.
The roles of fluctuations, dissipation, and the biharmonic force on the ratchet current were investigated in classical and quantum mechanical cases
by studying the dependence of the ratchet current on these parameters. 

In the classical case, we found that over the temperature range we studied,
the current realizes a maximum at an intermediate value,
and falls off for both high and low temperatures.
This decrease on the high temperature side is a result of thermal activation,
which induces dispersion of the distribution over the barrier in the high temperature region.
It the quantum mechanical case,
that tunneling enhances the current in the high barrier case,
while it suppresses the current in the low barrier case.
This is because the effect of the ratchet potential is weak in the case of a low
barrier, due to the dispersion of the distribution that arises from tunneling.
Moreover, in the quantum mechanical case, there is a classical Smoluchowski-like regime induced by tunneling effects. 
We also found that the phase lag of the biharmonic force is related to the response time of the system to the external perturbation. 

As we demonstrated, the dynamical behavior of the system is clearly and readily elucidated
by the time evolution of the Wigner distribution function.
Although we must introduce hierarchies of Wigner functions described by a discretized mesh,
a modern personal computer is sufficiently powerful to solve the equations of motion in a reasonable amount of time.
Numerical techniques, such as the optimization of the hierarchy,\cite{Shi09,YanPade10A,YanPade10B,Aspru11}
the utilization of a graphic processing unit (GPU),\cite{GPU11}
and memory allocation for massive parallel computing\cite{Schuten12} have been developed for the HEOM approach
in order to accelerate numerical calculations.
As an approximated method, it may be possible to adapt a multi-configuration time-dependent Hartree (MCTDH) approach,
which is shown to be an efficient method to solve Schr\"{o}dinger equations of motion.\cite{Meyer90,Thoss2003,Burghardt2008}
With these developments, the extension of the present study 
to the variety of multi-dimensional potential systems with non-Drude-type spectral distribution functions may be possible.

\begin{flushleft}
\Large{ \textbf{Acknowledgement}}
\end{flushleft}

YT is grateful for stimulating discussions with Professor Peter H\"{a}nggi and the hospitality of him and his group members during his stay at the University of Augsburg,
made possible by the Humboldt Foundation. AK acknowledges a research fellowship from Kyoto University. The authors are grateful for useful comments on the finite difference expressions of the HEOM with Dr. Atsunori Sakurai. This research is supported by a Grant-in-Aid for Scientific Research (B2235006) from the Japan Society for the Promotion of Science.

\pagebreak

\end{document}